\documentclass[conference]{IEEEtran}
\IEEEoverridecommandlockouts

\usepackage{amsmath,amssymb,amsfonts}
\usepackage{algorithmic}
\usepackage{graphicx}
\usepackage{textcomp}
\usepackage{xcolor}
\def\BibTeX{{\rm B\kern-.05em{\sc i\kern-.025em b}\kern-.08em
    T\kern-.1667em\lower.7ex\hbox{E}\kern-.125emX}}
\usepackage[style=trad-abbrv, backend=biber, maxnames=1, minnames=1, doi=false,isbn=false,url=false , giveninits=true, sortcites]{biblatex}
\usepackage{graphicx}
\usepackage{booktabs}
\usepackage{threeparttable}
\usepackage{hyperref}
\usepackage{cleveref}
\Crefname{figure}{Fig.}{Figs.}
\usepackage{siunitx}
\usepackage{amsthm}
\usepackage{caption}
\usepackage{enumitem}
\usepackage{subcaption}
\usepackage{float}
\usepackage{braket}
\usepackage{tikz}
\usepackage{yquant}
\useyquantlanguage{groups}
\yquantset{register/default name=}
\newtheorem{definition}{Definition}
\usepackage{mathtools}
\usepackage{stmaryrd}
\DeclarePairedDelimiter{\floor}{\lfloor}{\rfloor}
\newtheorem{example}{Example}

\usepackage{tabularx,environ}

\makeatletter
\newcommand{\problemtitle}[1]{\gdef\@problemtitle{#1}}%
\newcommand{\probleminput}[1]{\gdef\@probleminput{#1}}%
\newcommand{\problemquestion}[1]{\gdef\@problemquestion{#1}}%
\NewEnviron{problem}{
  \problemtitle{}\probleminput{}\problemquestion{}%
  \BODY%
  \par\addvspace{.5\baselineskip}
  \noindent
  \begin{tabularx}{.5\textwidth}{@{\hspace{\parindent}} l X c}
    \multicolumn{2}{@{\hspace{\parindent}}l}{\@problemtitle} \\
    \textbf{Input:} & \@probleminput \\
    \textbf{Problem:} & \@problemquestion%
  \end{tabularx}
  \par\addvspace{.5\baselineskip}
}
\makeatother
\begin{document}

\title{Deterministic Fault-Tolerant State Preparation for Near-Term Quantum Error Correction: \\Automatic Synthesis Using Boolean Satisfiability}

\author{
    \IEEEauthorblockN{
        Ludwig Schmid\IEEEauthorrefmark{1},
        Tom Peham\IEEEauthorrefmark{1},
        Lucas Berent\IEEEauthorrefmark{1},
        Markus Müller\IEEEauthorrefmark{2}\IEEEauthorrefmark{3},
        and Robert Wille\IEEEauthorrefmark{1}\IEEEauthorrefmark{4}
    }
    \IEEEauthorblockA{\IEEEauthorrefmark{1}%
    Chair for Design Automation,
        Technical University of Munich,
        80333 Munich, Germany
    }
    \IEEEauthorblockA{\IEEEauthorrefmark{2}%
      Institute for Quantum Information,
      RWTH Aachen University,
      52056 Aachen, Germany
    }
    \IEEEauthorblockA{\IEEEauthorrefmark{3}%
      Peter Grünberg Institute, Theoretical Nanoelectronics,
      Forschungszentrum Jülich,
      52425 Jülich, Germany
    }
    \IEEEauthorblockA{\IEEEauthorrefmark{4}%
    Software Competence Center Hagenberg GmbH,
        4232 Hagenberg, Austria
    }
}

\maketitle

\begin{abstract}
    To ensure resilience against the unavoidable noise in quantum computers, quantum information needs to be encoded using an error-correcting code, and circuits must have a particular structure to be fault-tolerant.
    Compilation of \mbox{fault-tolerant} quantum circuits is thus inherently different from the \mbox{non-fault-tolerant} case.
    However, automated fault-tolerant compilation methods are widely underexplored, and most known constructions are obtained manually for specific codes only.
    In this work, we focus on the problem of \emph{automatically} synthesizing fault-tolerant circuits for the deterministic initialization of an encoded state for a broad class of quantum codes that are realizable on current and near-term hardware. 
    
    To this end, we utilize methods based on techniques from classical circuit design, such as satisfiability solving, resulting in tools for the synthesis of (optimal) fault-tolerant state preparation circuits for near-term quantum codes.
    We demonstrate the correct fault-tolerant behavior of the synthesized circuits using \mbox{circuit-level} noise simulations.
    We provide all routines as \mbox{open-source} software as part of the \emph{Munich Quantum Toolkit}~(MQT) at \href{https://github.com/cda-tum/mqt-qecc}{https://github.com/cda-tum/mqt-qecc}.
\end{abstract}


\section{Introduction}\label{sec:intro}
A major difficulty in building a large-scale quantum computer is the fact that quantum bits and operations are inherently susceptible to errors. 
This high error liability leads to errors accumulating during the execution of a quantum algorithm, effectively rendering the results useless.
To amend this problem, active error correction and \mbox{fault-tolerance} are needed.

The key idea of a fault-tolerant quantum algorithm is to use an error-correcting code to encode quantum information using additional resources in a redundant manner to ensure resilience to errors~\cite{preskillReliableQuantumComputers1998,shorFaulttolerantQuantumComputation1996,kitaevQuantumComputationsAlgorithms1997,aharonovFaulttolerantQuantumComputation1997,nielsenQuantumComputationQuantum2010}. 
The goal is to control the spreading of errors between qubits in the system and, by that, avoid the uncontrollable accumulation of errors in the encoded information.

In a circuit-centric model of computation, the design of fault-tolerant quantum circuits encompasses many problems that resemble those in the domain of classical circuit design, such as compilation and synthesis tasks.
In general, however, the area of fault-tolerant quantum circuit synthesis is widely underexplored.
In this work, we focus on a particular instance of a circuit synthesis problem for quantum codes that are realizable on current and near-term hardware and that is centered around an essential step in fault-tolerant quantum computing: the initialization of an encoded state.

In general, synthesizing an optimal circuit whose output is an encoded state of a given code is highly non-trivial, as one must ensure that the process of obtaining the encoded state is fault-tolerant itself.
That is, it must be ensured that the encoded state does not contain too many errors on initialization~\cite{postlerDemonstrationFaulttolerantUniversal2022, peham2024automated, derksDesigningFaulttolerantCircuits2024, gotoMinimizingResourceOverheads2016}.
This preparation step can account for a significant part of the qubit and gate overhead~\cite{heussenEfficientFaulttolerantCode2024}.
For an important class of codes, called \emph{Calderbank-Shor-Steane} (CSS) codes~\cite{calderbankGoodQuantumErrorcorrecting1996, shorFaulttolerantQuantumComputation1996, steaneErrorCorrectingCodes1996}, a way of preparing an encoded state is to conduct specific measurements repeatedly, together with a correction procedure to produce the desired state~\cite{nielsenQuantumComputationQuantum2010}.
However, this method can be costly for near-term hardware.

Accordingly, promising state-of-the-art approaches for currently available hardware are centered around methods that rely on repeat-until-success schemes referred to as \mbox{\emph{non-deterministic}}~\cite{postlerDemonstrationFaulttolerantUniversal2022, heussenStrategiesPracticalAdvantage2023, ryan-andersonRealizationRealtimeFaulttolerant2021} to construct smaller circuits.
More precisely, they employ a (generally not fault-tolerant) \emph{state preparation circuit} first, followed by a \emph{verification circuit} to detect problematic errors. 
In case an error is detected, the protocol is started anew, and the overall circuit is run until a state in which no problematic errors are detected can be constructed.
The aim is to construct a circuit in this heralded fashion, such that, after a certain number of runs, an error-free state can be initialized with high probability.

An obvious drawback of this approach is its stochastic nature. 
That is, it might take many repetitions of unknown number to prepare the desired state successfully, which can be a critical factor in experiments~\cite{heussenStrategiesPracticalAdvantage2023}.
Recently, a method to eliminate the non-determinism in such a scheme was proposed for a particular error-correcting code instance~\cite{heussenStrategiesPracticalAdvantage2023}.
Here, the main idea is to extend the non-deterministic circuit by additional operations, we term \emph{correction circuit}, which allow one to infer which error most likely happened in case an error was detected in the verification part and then to use the information to convert it into an unproblematic error applying a suitable recovery operation.

From a circuit synthesis point of view, the downsides of the current state of the art with respect to this approach are that there are no optimality guarantees and that the construction is done mostly manually for specific codes~\cite{postlerDemonstrationFaulttolerantUniversal2022, heussenStrategiesPracticalAdvantage2023}. 
Hence, automated methods similar to \cite{peham2024automated,shuttyDecodingMergedColorSurface2022} that allow obtaining optimality guarantees are needed and can be used to apply the scheme to other codes without further manual inspection of the code at hand.

In this work, we apply techniques from classical circuit design, such as \emph{satisfiability solving} (SAT), to tackle the problem of synthesizing fault-tolerant state preparation circuits for small quantum codes that are implementable on current and near-term hardware. 
We propose automated methods applicable to a broad class of codes that allow us to obtain verification-, and correction circuits that are optimal in size. 

To foster research in the area of fault-tolerant circuit synthesis, we provide the resulting tools as open-source software as part of the \emph{Munich Quantum Toolkit}~(MQT)~\cite{willeMQTHandbookSummary2024}.

\section{Background}\label{sec:background}

In this section, we review the technical background regarding the spreading and correction of errors in quantum circuits, needed to motivate the considered problem in the following. %

Although qubit errors can be arbitrary \emph{unitary} operators, any such error can be written as a linear combination of the \mbox{so-called} \emph{Pauli matrices}, below, forming the multiplicative {Pauli group}:
\vspace{-1mm}
$$
I = \begin{bmatrix}
    1 & 0 \\ 0 & 1
\end{bmatrix}, X = \begin{bmatrix}
    0 & 1 \\ 1 & 0
\end{bmatrix}, Z = \begin{bmatrix}
    1 & 0 \\ 0 & -1
\end{bmatrix}, Y = i\begin{bmatrix}
    0 & -1 \\ 1 & 0
\end{bmatrix}.
$$

The Pauli group on $n$ qubits $\mathcal{P}_n$ is given by all combinations of Pauli operators acting on each qubit individually. 
An example is $O=I\otimes X \otimes Z =X_2Z_3 \in \mathcal{P}_3$ with the index indicating the qubit the operator is acting on non-trivially.
The number of such non-trivial terms is called the \emph{weight} of a Pauli operator $\mathrm{wt}(O)$.

\begin{figure}
     \vspace{-0.6cm}
     \centering
     \includegraphics[width=0.8\linewidth]{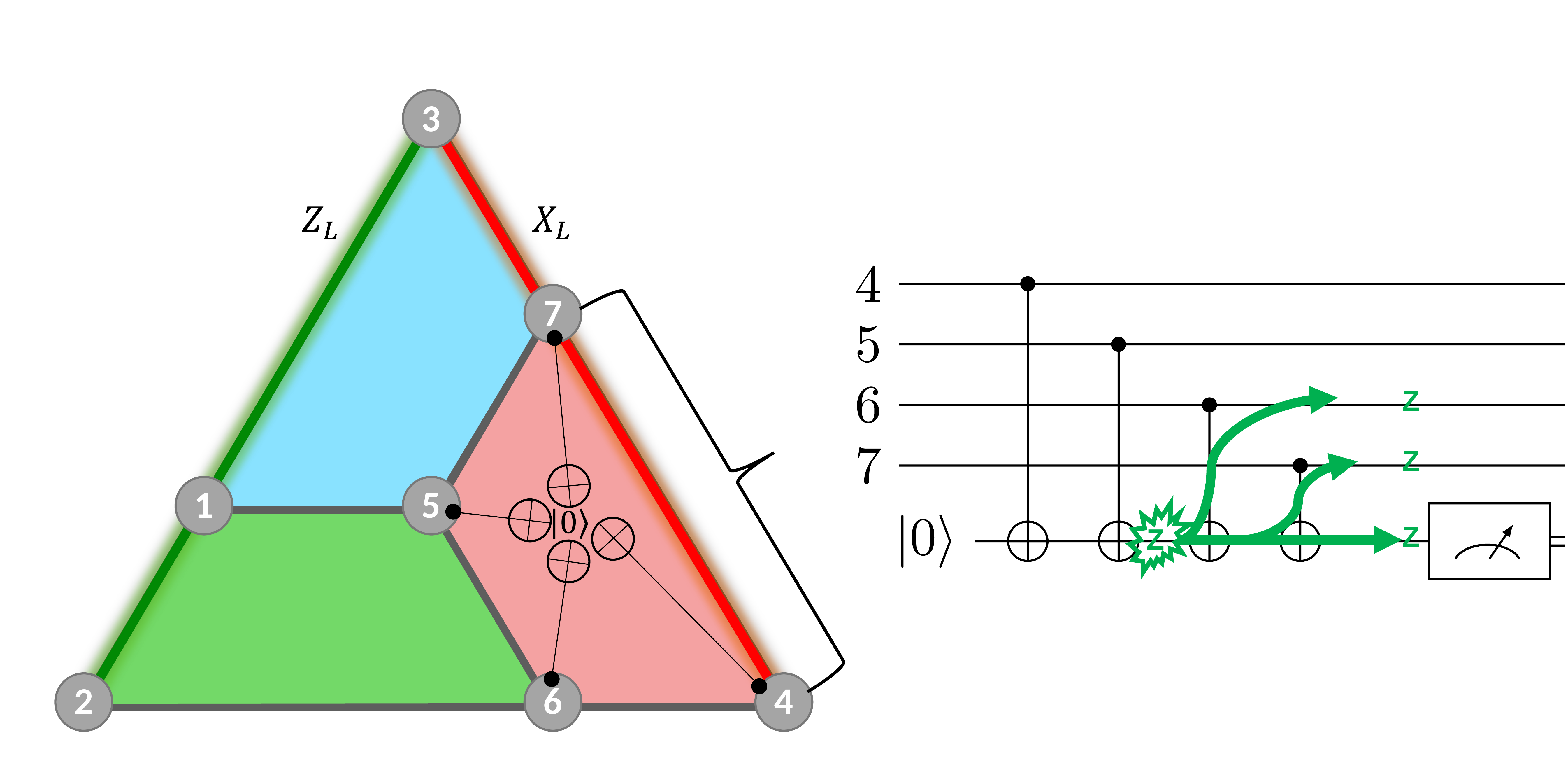}
     \caption{Steane code with a hook error on a $Z$ stabilizer.}
     \label{fig:steane}
 \end{figure}

How these Pauli operators propagate through circuits, especially through two-qubit gates, is essential in the design of fault-tolerant quantum circuits. 
For example, the circuit in~\Cref{fig:steane} illustrates how Pauli Zs propagate through the target of a CNOT. 
Analogously, Pauli Xs propagate through the control of a CNOT.

Quantum error-correcting codes encode information redundantly, representing encoded \emph{logical qubits} using multiple noisy \emph{physical qubits}. 
The logical state space of $k$ logical qubits encoded into $n$ physical qubits is then a $2^k$-dimensional subspace of $\mathcal{H}^{\otimes n}$.
A code with $n$ physical qubits encoding $k$ logical qubits with distance $d$ is denoted as $[[n, k, d]]$-code.

Stabilizer codes~\cite{gottesmanStabilizerCodesQuantum1997} are particularly important as they encompass most practically relevant codes.
Stabilizer codes are defined as the joint $+1$ eigenspace of a \emph{stabilizer group}~$S$, a subgroup of the Pauli group with $-I_n \notin S$. 
The elements of $S$ are called \emph{stabilizers}, and the codes are given by a minimal set of \emph{stabilizer generators} $S_1, \cdots, S_m$ such that $\langle S_1, \cdots, S_m \rangle = S$, where $m=n-k$.
The logical operators $X_L, Z_L$ are the operators that commute with the generators but are not in $S$ and 
the \emph{code distance}~$d$ of a stabilizer code is the weight of a minimal logical operator.
When talking about errors in the context of stabilizer codes, we are usually only interested in \emph{minimal-weight stabilizer-equivalent} representatives since we can multiply an error by a stabilizer without changing its effect. 
Therefore, for $e \in \mathcal{P}_n$, we define
$\mathrm{wt}_S(e)=\min_{e'\in \{s e\mid s\in S\}}wt(e').$

When all stabilizers of a code are either all $X$ or all $Z$, then it is a \emph{Calderbank-Shor-Steane} (CSS) code~\cite{calderbankGoodQuantumErrorcorrecting1996,steaneErrorCorrectingCodes1996, shorFaulttolerantQuantumComputation1996}, which is the class of codes we are going to focus on in this work.
\begin{example}\label{ex:steane}
The \emph{Steane} code is a $[[7, 1, 3]]$ CSS~\cite{bombinIntroductionTopologicalQuantum2013} code defined by the stabilizer generators $S = \{X_1X_2X_5X_6, X_1X_3X_5X_7, X_4X_5X_6X_7, Z_1Z_2Z_5Z_6, \\ Z_1Z_3Z_5Z_7, Z_4Z_5Z_6Z_7\}$. 
The Steane code can be illustrated as a tiling of a triangle as in~\Cref{fig:steane} with qubits on vertices and stabilizer generators on the faces. Minimum-weight logical operators are defined along the sides of the triangle: $X_L=X_3X_4X_7$ and $Z_L=Z_1Z_2Z_3$.
\end{example}

Since the code space is the common $+1$ eigenspace of all stabilizers, measuring the stabilizer generators gives a way of extracting information about errors that happened on an encoded state without disturbing the logical state. 
A \mbox{$Z$-type} stabilizer in a CSS code can be measured by preparing an ancillary qubit in the $\ket{0}$ state, entangling it with the \mbox{non-trivial} qubits of the stabilizer by performing CNOTs with the ancilla as the target and measuring the ancilla.
A measurement of $\ket{1}$ indicates the presence of an odd number of $X$ errors on the data qubits.
An example of such a measurement on the Steane code can be seen in~\Cref{fig:steane}, and similar circuits for arbitrary multi-qubit Pauli operators can be constructed.

Measurement information can be used to decode and correct errors. 
Unfortunately, using an error correction code does not by itself guarantee fault-tolerance. 
An important consideration is the propagation of faults when performing entangling operations on data qubits.
As illustrated in~\Cref{fig:steane-det}, an entangling gate can amplify a single-qubit error to a weight-two error.
This leads to the notion of strict fault-tolerance:

\begin{definition}[\textbf{Strict fault-tolerance}~\cite{paetznickFaulttolerantAncillaPreparation2013,derksDesigningFaulttolerantCircuits2024}]\label{def:ft}
  A circuit is \emph{strictly fault-tolerant} if for all $t \leq \floor{\frac{d}{2}}$, any error of probability order $t$ propagates to an error of weight at most $t$.
\end{definition}

\begin{example}
    Measuring a weight-four $Z$ stabilizer generator of the Steane code with the circuit in~\Cref{fig:steane} is not fault-tolerant, since a single $Z$ error on the ancilla between the second and third CNOT propagates to a weight-two error on the data qubits.
\end{example}

In particular, CNOTs between data qubits often violate the requirement of strict fault-tolerance since single-qubit errors on the data propagate further.
Therefore, circuits used to prepare logical states are themselves not necessarily \mbox{fault-tolerant}, and additional considerations need to be taken to avoid the uncontrolled spreading of errors.

\section{Motivation and Problem Description}\label{sec:motiv-probl-descr}

In this section, we provide an illustrative overview of the synthesis of deterministic fault-tolerant state preparation circuits and motivate the proposed method.

\subsection{Non-Deterministic Fault-Tolerant State Preparation}

Due to error propagation along entangling gates, naive synthesis of state preparation circuits is not fault-tolerant in general. 
Intuitively, this means that Pauli errors may propagate through the circuit, e.g., through CNOT gates, which may cause errors on a few qubits to spread to higher-weight errors, violating \Cref{def:ft}. 

\begin{figure}[t]
  \centering
    \captionsetup{
    labelfont=bf,        %
    justification=raggedright,
}
\includegraphics[width=\linewidth]{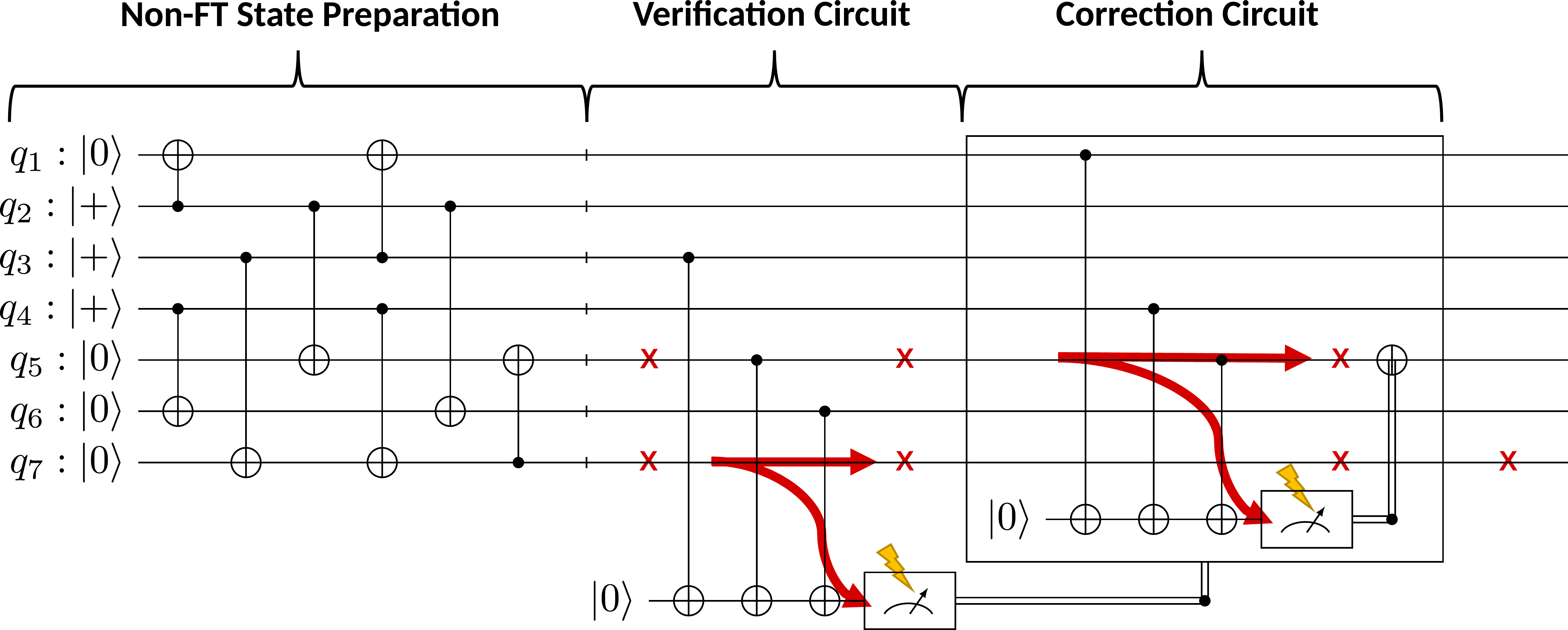}
  \caption{Deterministic fault-tolerant state preparation of the Steane code $\ket{0}_L$ with error propagation and correction.}
  \label{fig:steane-det}
\end{figure}

\begin{example}
\label{ex:steane-non-ft}
Consider the first part of the circuit shown in~\Cref{fig:steane-det}, which prepares the logical $\ket{0}_L$ state of the Steane code using a unitary encoding circuit. %
    This circuit is not fault-tolerant by~\Cref{def:ft}, since an $X$ error on the control qubit before the last CNOT gate propagates to a weight-two error on the final state.
\end{example}

A possible approach to synthesizing state preparation circuits that are fault-tolerant is commonly termed \mbox{\emph{non-deterministic}} and essentially relies on a \mbox{repeat-until-success-protocol} \cite{buttFaultTolerantCodeSwitchingProtocols2024, bermudezFaulttolerantProtectionNearterm2019,heussenMeasurementFreeFaultTolerantQuantum2024, postlerDemonstrationFaulttolerantUniversal2022,ryan-andersonRealizationRealtimeFaulttolerant2021, bluvsteinLogicalQuantumProcessor2024, m.p.dasilvaDemonstrationLogicalQubits2024, pogorelovExperimentalFaulttolerantCode2024}.
Therein, the overall state preparation circuit consists of two parts: 1) a circuit that prepares the desired logical state and that is generally not fault-tolerant and 2) a \emph{verification circuit} that is synthesized depending on the structure of the preparation circuit and detects dangerous spreading errors. 
If the verification circuit does not detect any dangerous spreading errors, the state is guaranteed to not have too many errors according to~\Cref{def:ft}.

\begin{example}
    Continuing the previous example, the first measurement in~\Cref{fig:steane-det} detects any propagated problematic \mbox{two-qubit} error as illustrated by the propagation of the error onto the measurement ancilla. 
    Repeating the state preparation until this measurement returns $+1$ prepares the state $\ket{0}_L$ and ensures that an uncorrectable error occurs only with probability $O(p^2)$ where $p$ is the physical error rate.
\end{example}

While this scheme produces logical states with the desired logical error rate, the obvious drawback to this approach is the need for potentially multiple rounds of state preparation before a state passes verification.
This may pose issues in experiments, e.g., regarding synchronization when a state needs to be ready at a certain time~\cite{heussenStrategiesPracticalAdvantage2023}.

\subsection{Deterministic Fault-Tolerant State Preparation}
\label{sec:determ-fault-toler}

Recently, in Ref.~\cite{heussenStrategiesPracticalAdvantage2023}, the authors proposed a way to circumvent this problem by extending the synthesized circuit with a \emph{correction circuit}, which consists of additional measurements and a conditional recovery operation. 
Intuitively, the idea is that this circuit is used to determine which error happened in case the verification indicates a potentially dangerous error. 
The additional measurements allow for deriving a recovery operation that either directly corrects the error or converts it to a correctable one. 
Thus, the state does not need to be discarded.
This removes the need for post-selection and thus renders the protocol \emph{deterministic}.

\begin{figure*}[t]
    \centering
    \vspace{-0.8cm}
    \includegraphics[width=1\linewidth]{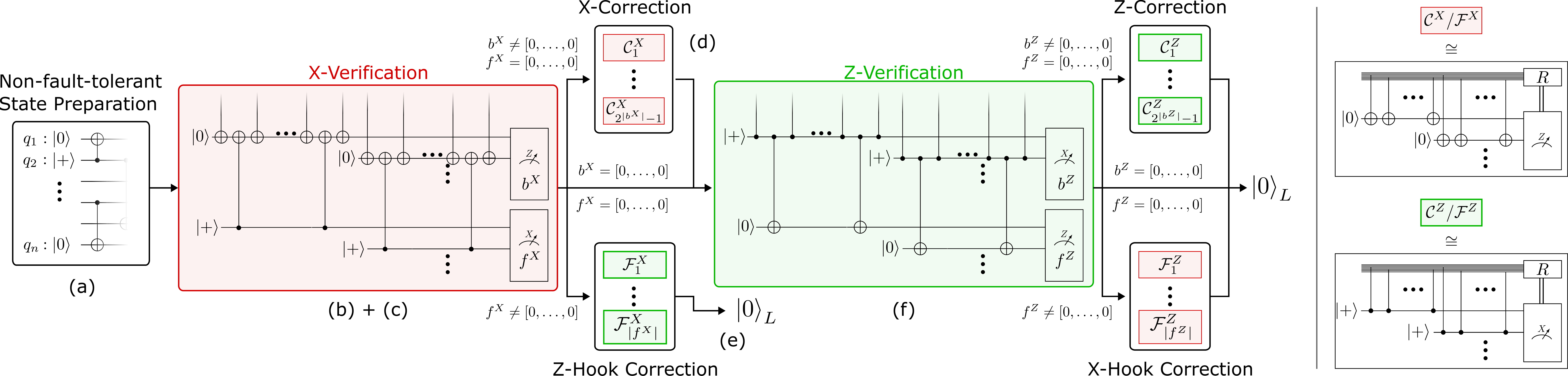}
    \caption{Deterministic fault-tolerant state preparation protocol for $d<5$ CSS codes with boxes illustrating the corresponding circuits. Information and data qubit flow are depicted as arrows. The circuits on the right illustrate the structure of additional measurements and recovery operations.
    \textbf{(a)} The state is unitarily prepared via a generally non-fault-tolerant circuit. 
    \textbf{(b)} Verification measurements are performed to check whether a problematic X error propagated in the preparation circuit.
    \textbf{(c)} The verification measurements are flagged to detect problematic hook errors.
    \textbf{(d)} Depending on the verification measurements~$b^{X}$, a suitable correction circuit is executed.
    \textbf{(e)} If a hook error is detected on~$f^{X}$, no error happened during the state preparation circuit, so the protocol terminates after the appropriate correction.
    \textbf{(f)} If no hook error is detected, the same verification and correction procedure is performed for Z errors. 
    At the end of the protocol, the logical state is prepared deterministically.}
    \label{fig:full-scheme}
\end{figure*}

\begin{example}
    Conditioned on the measurement in ~\Cref{fig:steane-det}, the circuit in the box is executed if the first measurement yields a $-1$. The second measurement gives enough information to correct any propagated error such that an error with at most weight one is left at the end of the circuit. This fulfills the fault-tolerance requirement of~\Cref{def:ft} without requiring a restart of the logical encoding protocol.
\end{example}

A downside of the current state-of-the-art, such as proposed in Ref.~\cite{heussenStrategiesPracticalAdvantage2023} is that the corresponding circuits are derived manually and that there are no optimality guarantees in terms of the size of the circuits.
Thus, the techniques are not generally applicable, and the construction of such deterministic circuits remains a tedious, case-by-case task.

In this work, we generalize the existing scheme of Ref.~\cite{heussenStrategiesPracticalAdvantage2023} and provide an automated method to synthesize such deterministic fault-tolerant state preparation circuits.
Given a \mbox{non-deterministic} fault-tolerant state preparation for an~$[[n,k,d]]$ CSS code with $d < 5$ and corresponding verification, we determine a correction circuit such that the overall procedure is fault-tolerant according to~\Cref{def:ft}.
The proposed technique relies on satisfiability-solving methods, which are commonly used in classical circuit design.
The synthesized correction circuits are guaranteed to be optimal in the sense that the additional measurements needed to determine a correction have minimal weight and, therefore, require as few CNOTs as possible.

In the following, we will describe this synthesis method in detail by showing how to encode the problem of correction circuit synthesis into a Boolean formula.

\section{Technical Implementation}\label{sec:techn-impl}

In this section, we detail the proposed fault-tolerant deterministic state preparation protocol.
\Cref{fig:full-scheme} shows a flow diagram illustrating the complete protocol, described in the following.
It consists of two layers corresponding to the verification and correction of $X$ and $Z$ errors, respectively.
Note that depending on the code at hand, parts of the protocol (up to whole layers) can be omitted.
Also, for part~\textbf{(a)} and~\textbf{(b)} of this protocol, already existing solutions~\cite{zenQuantumCircuitDiscovery2024,peham2024automated,shuttyDecodingMergedColorSurface2022} can be employed.

Given an $[[n,k,d<5]]$ CSS code with stabilizer group~$S$, a non-fault-tolerant state preparation circuit~$\mathcal{C}$ for a Pauli eigenstate of this code and an appropriate verification circuit~$\mathcal{V}$, the task is to find the correction circuit.
I.e., a set of measurements and Pauli recoveries such that if at most $1$ error occurred in the circuit, an error with weight at most $1$ remains on the output state.
Since we focus on CSS codes, we consider the verification for X and Z errors separately and thus construct the correction circuits $\mathcal{V}_X$ and $\mathcal{V}_Z$ independently. 

We define $\mathcal{E}_X(\mathcal{C})$ ($\mathcal{E}_Z(\mathcal{C})$) as the set of all dangerous $X$~($Z$)~errors $e$ stemming from single faults in $\mathcal{C}$ with $\mathrm{wt}_S(e) \geq 2$.
We can interpret an $X$~($Z$) Pauli operator as a vector $p \in \mathbb{F}_2^n$ such that the $i$-th coordinate $p[i]=1$ if the operator acts on qubit $i$ and as identity on the other qubits $j\neq i$.
Then, the condition for $\mathcal{V}$ to be a valid verification circuit is that every error $e \in \mathcal{E}_X(\mathcal{C})$ ($\mathcal{E}_Z(\mathcal{C})$) anticommutes with at least one stabilizer $s$ measured in $\mathcal{V}$, i.e., $\langle e , s \rangle = 1$, where $\langle \cdot \mid \cdot \rangle$ is the standard inner product in $\mathbb{F}_2^n$.

The syndrome $b$ of an error $e$ is a vector $b \in \mathbb{F}_2^{|\mathcal{V}|}$ such that $b[i] = \langle e \mid m_i \rangle$, where $m_i$ is the $i$-th measurement in $\mathcal{V}$.
Syndromes do not correspond uniquely to errors, and thus, the possible syndromes partition the set of errors $\mathcal{E}_X(\mathcal{C})$~($\mathcal{E}_Z(\mathcal{C})$) into sets~$\mathcal{E}_b$ that have the same syndrome.
The problem of finding a correction circuit can be considered for each $\mathcal{E}_b$ independently, significantly reducing the complexity.

A correction $c$ for an error $e$ is a Pauli operator such that $\mathrm{wt}_S(e+c) \leq 1$ and can also be represented as a vector in $\mathbb{F}_2^n$. 
Note that a single $c$ might correct multiple errors.
Therefore, we can associate~$c$ with its correction set $F(c) \subseteq \mathcal{E}_b$, which are all the errors in $\mathcal{E}_b$ corrected by~$c$.
In general, we need to measure multiple additional stabilizers to further partition~$\mathcal{E}_b$ until all errors within each set are correctable by the same~$c_b$.
Then, the corresponding correction can be applied as a recovery operation whenever $b$ is measured.

Regarding single-qubit errors, additional care has to be taken.
Although not a problem in and of themselves, there might be single-qubit errors with non-trivial syndrome~$b$ (e.g.,~a measurement error).
When applying the appropriate correction $c_b$ after the additional measurements, the product of the \mbox{single-qubit} error and the corrections should also have weight at most one.
This can be ensured by additionally including these \mbox{single-qubit} errors in the error set $\mathcal{E}_b$.

Due to the high error rates on near-term quantum computing devices, it is natural to aim to synthesize correction circuits with as few measurements and CNOT gates as possible.
To this end, we can frame the problem of correction circuit synthesis as a decision problem to solve it optimally, employing satisfiability solvers such as Z3~\cite{demouraZ3EfficientSMT2008}. Satisfiability solvers have demonstrated their utility in Clifford circuit synthesis~\cite{shuttyDecodingMergedColorSurface2022,peham2024automated,pehamDepthOptimalSynthesisClifford2023}, providing optimal solutions but exhibiting poor scalability.

\begin{problem}
\problemtitle{\textsc{Correction Circuit Synthesis}}
    \probleminput{$X$ ($Z$) stabilizer generators $S_X$ ($S_Z$), a set of errors $\mathcal{E}$ and integers $u$, $v$}
    \problemquestion{For given errors $\mathcal{E}$ is there a set of stabilizers $s_1, \cdots, s_u \in \langle S_X \rangle$ ($\langle S_Z \rangle$) with bounded weight \mbox{$\sum_{i=0}^u \mathrm{wt}(s_i) \leq v$} such that all errors with the same syndrome $b$ are reduced to a correctable error ($\mathrm{wt}_S(e) \leq 1, e \in \mathcal{E}$) by the same recovery $c_b$.}
\end{problem}

By encoding this task as a Boolean satisfiability problem, we can obtain a solution that is optimal in the number of measurements (CNOTs) by iterating over the number of generators and weight parameters $u$ and $v$, respectively, until we find $u$ and $v$ such that \textsc{Correction Circuit Synthesis} is a \textsc{yes}-instance for $u$ and $v$ but not for $u-1$ (or $v-1$).

We again use the correspondence to $\mathbb{F}_2$ to encode generators, errors, and stabilizers as binary variables.
Hence, it is straightforward to encode the correction circuit synthesis as a Boolean formula using suitable constraints and binary operations.
The exact formulation of the constraints are given in the source code at \href{https://github.com/cda-tum/mqt-qecc}{https://github.com/cda-tum/mqt-qecc}.

If the formula is satisfiable, the stabilizer measurements and the correction can be obtained from the assignment found by the solver.
This way, we can synthesize optimal circuits for the boxes of step \textbf{(d)} of the protocol in \Cref{fig:full-scheme}.

Until now, we assumed that all problematic errors stem from propagated errors in the state preparation circuit itself. 
But as shown in \Cref{fig:steane}, verification measurements can introduce so-called \emph{hook errors} that spread from the ancilla onto the data qubits.
These hook errors can be detected by measuring the stabilizer \emph{flag-fault-tolerantly}~\cite{chamberlandFlagFaulttolerantError2018} by adding a single additional \emph{flag} ancilla and two CNOT gates that herald the occurrence of such a hook error.
While this flag-fault-tolerance scheme is non-deterministic in general, we can make it deterministic by combining the flag qubit syndromes $f$ with the other measurement syndrome bits $b$ before synthesizing the correction circuits.
Note that occasionally, it might be preferable not to flag certain stabilizer measurements if the corresponding hook errors are not dangerous or equivalent to one of the regular spreading errors.
We see examples for this case in \Cref{sec:evaluation}.

Since we aim to protect the state preparation circuit against single errors (due to our assumption $d<5$), 
the occurrence of a hook error and an error in the state preparation circuit are mutually exclusive. 
Therefore, we can consider the correction circuit synthesis for hook errors independently from other errors, reducing complexity. 
Moreover, if the hook error was detected in the first verification layer and we successfully applied the correction circuit, we do not have to run the second verification layer as shown in Step \textbf{(e)} of \Cref{fig:full-scheme}.

Clearly, the correction circuits depend on the preceding verification circuit, and in general, there are multiple equivalent verification circuits that lead to significantly different correction circuits.
Thus, to further optimize the overall circuit one can try to optimize both parts---verification and correction circuit---jointly. 
Since we are looking at codes with $d<5$, however, the relevant state preparation circuits are quite small and have few spreading errors, leading to a small number of equivalent verification circuits.
Consequently, we can synthesize all minimal verification circuits using existing optimal synthesis techniques~\cite{peham2024automated} and synthesize corrections for each of them.
We refer to this as the \emph{global optimization procedure}.

\section{Evaluation}\label{sec:evaluation}
To evaluate the proposed method, we generate deterministic fault-tolerant state preparation circuits for $\ket{0}_\mathrm{L}$ of various \mbox{$d<5$} CSS codes.
We employ the method described above to synthesize optimal correction circuits and run the global optimization scheme to explore potentially better verification and correction circuits.

First, we focus on the metrics of the generated circuits, including the number of measurements and the number of CNOT gates.
Second, we simulate the generated circuits under circuit-level noise and show their \mbox{fault-tolerance} numerically.

\subsection{Circuit Generation}\label{sec:circuit-generation}

\begin{table*}[t]
  \centering
    \caption{Circuit metrics for synthesizing deterministic fault-tolerant state preparation circuits for $\ket{0}_{\mathrm{L}}$ of various codes.}\label{tab:circuits_metrics}
    \resizebox{\textwidth}{!}{
      \begin{threeparttable}

\begin{tabular}{@{}llcc|ccS[table-format=2.0]ccccc|cccccccc|cS[table-format=2.0]SS@{}}
\toprule
 &
  \multicolumn{1}{c}{} &
   &
   &
  \multicolumn{8}{c|}{First Layer} &
  \multicolumn{8}{c|}{Second Layer} &
  \multicolumn{4}{c}{Total} \\
 &
   &
   &
   &
  \multicolumn{4}{c}{Verification} &
  \multicolumn{4}{c|}{Correction} &
  \multicolumn{4}{c}{Verification} &
  \multicolumn{4}{c|}{Correction} &
  \multicolumn{2}{c}{Verification} &
  \multicolumn{2}{c}{Correction} \\ \cmidrule(l){5-24} 
 &
   &
   &
   &
  \multicolumn{2}{c}{\#ANC} &
  \multicolumn{2}{c}{\#CNOT} &
  \multicolumn{2}{c}{\#ANC} &
  \multicolumn{2}{c|}{\#CNOT} &
  \multicolumn{2}{c}{\#ANC} &
  \multicolumn{2}{c}{\#CNOT} &
  \multicolumn{2}{c}{\#ANC} &
  \multicolumn{2}{c|}{\#CNOT} &
  {$\Sigma\,$ANC} &
  {$\Sigma\,$CNOT} &
  {$\varnothing\,$ANC} &
  {$\varnothing\,$CNOT} \\
Codename &
  \multicolumn{1}{l|}{$\llbracket n, k, d \rrbracket$} &
  State Prep~$^{*}$ &
  Verif.~$^{**}$ &
  $a_m$ &
  $a_f$ &
  $w_m$ &
  $w_f$ &
  $a_m$ &
  $a_f$ &
  $w_m$ &
  $w_f$ &
  $a_m$ &
  $a_f$ &
  $w_m$ &
  $w_f$ &
  $a_m$ &
  $a_f$ &
  $w_m$ &
  $w_f$ &
   &
   &
   &
   \\ \midrule
Steane~\cite{bombinIntroductionTopologicalQuantum2013} &
  \multicolumn{1}{l|}{$\llbracket 7, 1, 3 \rrbracket$} &
  Opt/Heu &
  Opt/Global &
  1 &
   &
  3 &
   &
  [1] &
   &
  \textbf{[3]} &
   &
   &
   &
   &
   &
   &
   &
   &
   &
  1 &
  3 &
  1 &
  3 \\
Shor~\cite{shorSchemeReducingDecoherence1995} &
  \multicolumn{1}{l|}{$\llbracket 9, 1, 3 \rrbracket$} &
  Heu &
  Opt &
  2 &
  1 &
  5 &
  2 &
  [1,1,0] &
  [0] &
  [3,3,0] &
  [0] &
   &
   &
   &
   &
   &
   &
   &
   &
  3 &
  7 &
  0.5 &
  1.5 \\
 &
  \multicolumn{1}{l|}{} &
  Heu &
  Global &
  2 &
  1 &
  5 &
  2 &
  \textbf{[1,0,0]} &
  \textbf{[0]} &
  \textbf{[3,0,0]} &
  \textbf{[0]} &
   &
   &
   &
   &
   &
   &
   &
   &
  3 &
  7 &
  0.25 &
  0.75 \\
 &
  \multicolumn{1}{l|}{} &
  Opt &
  Opt/Global &
  1 &
  1 &
  3 &
  2 &
  \textbf{[1]} &
  \textbf{[0]} &
  \textbf{[3]} &
  \textbf{[0]} &
   &
   &
   &
   &
   &
   &
   &
   &
  2 &
  5 &
  0.5 &
  1.5 \\
Surface~\cite{bombinOptimalResourcesTopological2007} &
  \multicolumn{1}{l|}{$\llbracket 9, 1, 3 \rrbracket$} &
  Opt/Heu &
  Opt/Global &
  1 &
  1 &
  3 &
  2 &
  \textbf{[1]} &
  \textbf{[0]} &
  \textbf{[3]} &
  \textbf{[0]} &
   &
   &
   &
   &
  \textbf{} &
  \textbf{} &
  \textbf{} &
  \textbf{} &
  2 &
  5 &
  2 &
  3 \\
Ref.~\cite{QECCBoundsCircuits} &
  \multicolumn{1}{l|}{$\llbracket 11, 1, 3 \rrbracket$} &
  Heu &
  Opt &
  2 &
   &
  8 &
  \textbf{} &
  [1,1,1] &
   &
  [6,3,3] &
  \textbf{} &
  1 &
  1 &
  4 &
  2 &
  [1] &
  [1] &
  [4] &
  [3] &
  3 &
  12 &
  1 &
  6.5 \\
 &
  \multicolumn{1}{l|}{} &
  Heu &
  Global &
  2 &
   &
  8 &
   &
  \textbf{[1,1,0]} &
  \textbf{} &
  \textbf{[3,5,0]} &
  \textbf{} &
  1 &
  1 &
  4 &
  2 &
  \textbf{[1]} &
  \textbf{[1]} &
  \textbf{[4]} &
  \textbf{[3]} &
  3 &
  12 &
  0.8 &
  3 \\
Tetrahedal~\cite{steaneQuantumReedMullerCodes1999} &
  \multicolumn{1}{l|}{$\llbracket 15, 1, 3 \rrbracket$} &
  Opt/Heu &
  Opt/Global &
  1 &
  1 &
  3 &
  2 &
  \textbf{[2]} &
  \textbf{[0]} &
  \textbf{[6]} &
  \textbf{[0]} &
   &
   &
   &
   &
  \textbf{} &
  \textbf{} &
  \textbf{} &
  \textbf{} &
  2 &
  5 &
  1 &
  3 \\
Hamming~\cite{steaneSimpleQuantumErrorcorrecting1996} &
  \multicolumn{1}{l|}{$\llbracket 15, 7, 3 \rrbracket$} &
  Heu &
  Opt/Global &
  2 &
   &
  7 &
   &
  \textbf{[2,2,2]} &
  \textbf{} &
  \textbf{[6,6,6]} &
  \textbf{} &
   &
   &
   &
   &
  \textbf{} &
  \textbf{} &
  \textbf{} &
  \textbf{} &
  2 &
  6 &
  2 &
  6 \\
 &
  \multicolumn{1}{l|}{} &
  Opt &
  Opt/Global &
  2 &
   &
  6 &
   &
  \textbf{[2,2,0]} &
  \textbf{} &
  \textbf{[6,6,0]} &
  \textbf{} &
   &
   &
   &
   &
  \textbf{} &
  \textbf{} &
  \textbf{} &
  \textbf{} &
  2 &
  6 &
  1.67 &
  4 \\
Carbon~\cite{m.p.dasilvaDemonstrationLogicalQubits2024} &
  \multicolumn{1}{l|}{$\llbracket 12, 2, 4 \rrbracket$} &
  Opt &
  Opt/Global &
  1 &
   &
  6 &
   &
  \textbf{[3]} &
  \textbf{} &
  \textbf{[12]} &
  \textbf{} &
  1 &
  1 &
  6 &
  2 &
  \textbf{[1]} &
  \textbf{[1]} &
  \textbf{[6]} &
  \textbf{[4]} &
  3 &
  14 &
  1.67 &
  7.33 \\
 &
  \multicolumn{1}{l|}{} &
  Heu &
  Opt &
  2 &
   &
  8 &
   &
  \textbf{[2,1,1]} &
  \textbf{} &
  \textbf{[8,4,4]} &
  \textbf{} &
  1 &
  1 &
  6 &
  2 &
  \textbf{[2]} &
  \textbf{[1]} &
  \textbf{[8]} &
  \textbf{[4]} &
  4 &
  16 &
  1.4 &
  5.6 \\
Ref.~\cite{QECCBoundsCircuits} &
  \multicolumn{1}{l|}{$\llbracket 16, 2, 4 \rrbracket$} &
  Heu &
  Opt &
  2 &
   &
  8 &
   &
  \textbf{[2,2,0]} &
  \textbf{} &
  \textbf{[8,8,0]} &
  \textbf{} &
  2 &
  2 &
  12 &
  4 &
  \textbf{[2,0,1]} &
  \textbf{[1,2]} &
  \textbf{[12,0,4]} &
  \textbf{[4,8]} &
  6 &
  24 &
  1.25 &
  5.5 \\
Tesseract\cite{reichardtDemonstrationQuantumComputation2024} &
  \multicolumn{1}{l|}{$\llbracket 16, 6, 4 \rrbracket$} &
  Heu &
  Opt/Global &
  2 &
   &
  8 &
   &
  \textbf{[2,2,2]} &
  \textbf{} &
  \textbf{[8,8,8]} &
   &
  1 &
  1 &
  8 &
  2 &
  \textbf{[1]} &
  \textbf{[2]} &
  \textbf{[8]} &
  \textbf{[8]} &
  4 &
  18 &
  1.8 &
  8 \\ \bottomrule
\end{tabular}

        \begin{tablenotes}
        \item The number of necessary ancillae/measurements ($a$) and their corresponding summed weight/CNOT count ($w$) for the verification and the conditional corrections. Stabilizer measurements ($m$) and flag measurements ($f$) are listed separately and indicated by the subscript. For the correction, each number within the square brackets corresponds to one of the possible branches depending on the verification outcome. Entries in bold denote the current best metrics.
        \item $^{*}$ Synthesized using either the optimal (Opt) or heuristic (Heu) method from Ref.~\cite{peham2024automated}.
        \item $^{**}$ Synthesized using either the optimal (Opt) or globally optimized (Global) method described in \Cref{sec:techn-impl}.
      \end{tablenotes}
      \end{threeparttable}
      }
\end{table*}

\Cref{tab:circuits_metrics} provides an overview of the circuit metrics for the generated verification and correction circuits.
For various \mbox{$d<5$} CSS codes, we generate the state preparation circuits and the verification circuits using Ref.~\cite{peham2024automated} where the fourth column indicates which of the two synthesis methods, optimal (Opt) or heuristic (Heu), was used.
For the synthesis of the verification, we also apply the global optimization procedure (Global), which explores all possible verification and correction circuits, returning the overall best result.
If multiple methods are named in one line, all combinations result in the same circuit metrics (although generally not the same circuits).

For each layer and building block, we report the number of ancillae $a$ and the number of CNOT gates $w$ listed separately for additional measurements ($m$) and flags ($f$) indicated by the subscript. Verification measurements are executed during each run, while correction measurements are only executed conditionally if one of the verification or flag measurements is triggered (indicated by square brackets).  
The multiple numbers for the corrections correspond to the different branching possibilities based on the measurement outcome of the verification part, where for $a_m$ verification measurements, there are $2^{a_m}-1$ possible branches, excluding the trivial case where none of the measurements trigger.
A CNOT weight of $w_{m/f}=0$ indicates that no additional measurement needs to be performed for this branch because all problematic errors can be directly deterministically reduced to a correctable error.
The last column shows the metrics for the complete procedure, where for the verification, the sum of both layers is considered, as all measurements are executed every time.
For the correction circuits, the average over all branches is shown as an estimate of the expected cost per run.

The results show that for many codes, it is sufficient to employ only a single layer of verifying and correction.
For this, two conditions need to be fulfilled:
First, the state preparation only introduced problematic errors of one type (X or Z), which can then be checked within a single layer.
Second, by flagging some verification measurements, one can deterministically check and correct for problematic hook errors, omitting the need for a second verification layer.
Note that none of the flag corrections require additional measurements in the considered cases.

In general, this is not possible for larger codes, and a full second layer is required, as illustrated in \Cref{fig:full-scheme}.
Nevertheless, in some cases, it is possible to leave the first layer unflagged and capture the problematic hook errors entirely in the second layer with no or only moderate overhead.
This, clearly, does not work for the second layer, and the verification measurements need to be flagged to correct problematic hook errors deterministically.

The global optimization, exploring all possible verification and correction circuits, yields equivalently good circuits in most cases.
For the Carbon and the $\llbracket 16, 2, 4 \rrbracket$ code, the global optimization was canceled after two hours of runtime.
But for the Shor and the $\llbracket 11, 1, 3 \rrbracket$ code, it yields correction circuits with fewer ancillae and CNOTs, indicating the potential for improvement at the cost of computation time.

In summary, all numbers of the global optimization reported in this work represent the currently best-known circuits to our current knowledge, improving, in particular, on the hand-crafted solution of Ref.~\cite{heussenStrategiesPracticalAdvantage2023}.

\subsection{Simulation}\label{sec:simulation}

\begin{figure}
  \centering
  \vspace{-6mm}
  \includegraphics[width=.75\linewidth]{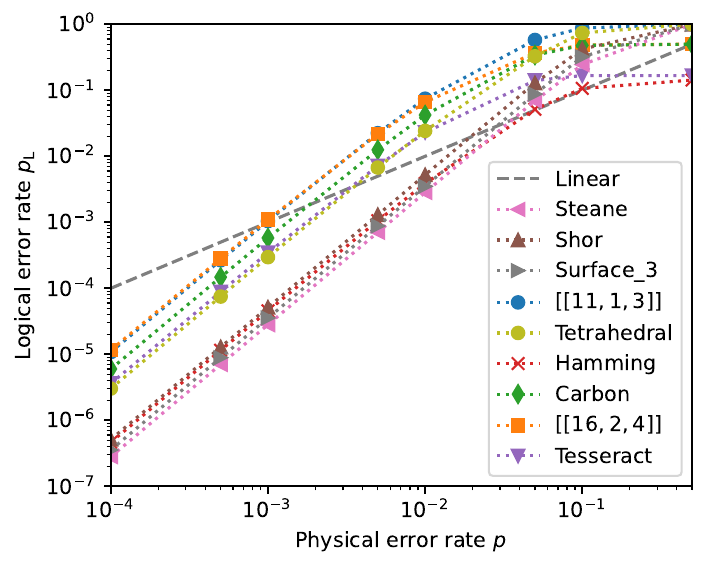}
   \vspace{-3mm}
  \caption{The logical error rates for the $\ket{0}_\mathrm{L}$ states of the heuristic state preparation and optimal verification from \Cref{tab:circuits_metrics}.}
  \vspace{-4mm}
  \label{fig:logical-error-rates}
\end{figure}

To show the correct fault-tolerant behavior of the synthesized circuits, we perform circuit-level noise simulations.
For this, we use \emph{Dynamic Subset Sampling} (DSS)~\cite{heussenDynamicalSubsetSampling2024} and the corresponding Python package Qsample~\cite{winterDpwinterQsample2024}.

We use the package internal \texttt{E1\_1} error model, corresponding to a standard depolarizing error model with a single physical error rate $p$ for single-qubit, two-qubit, and measurement errors.
The overall protocol is followed by a perfect round of error correction using lookup table decoding and, finally, measuring all data qubits destructively.
A logical error is registered if the resulting classical bitstring anticommutes with any of the logical operators of the Pauli eigenstate.
For each circuit, we sample 8000 runs at $ p_{\mathrm{max}}=0.1 $ and, then, employ DSS to compute the logical error rate for error \mbox{rates~$p < p_{\mathrm{max}}$}.

The resulting logical error rates are shown in~\Cref{fig:logical-error-rates}.
The results of the simulations show the expected quadratic scaling of the logical error rate, indicating that two independent physical errors are necessary to result in a logical error.
This is the desired behavior for a $d<5$ code, showing that a logical error occurs only with $O(p^2)$ and, therefore, numerically demonstrates the \mbox{fault-tolerance} of the circuits.

\section{Conclusion}\label{sec:conclusion}
In this work, we proposed a circuit design method to synthesize fault-tolerant deterministic state preparation circuits using satisfiability solvers for $d<5$ CSS code instances.
The proposed techniques can be used to extend already existing non-deterministic verification circuits by constructing and appending an additional correction circuit to avoid the stochastic repeat-until-success process in non-deterministic schemes.
Furthermore, we provide a global optimization procedure that synthesizes the verification and correction circuit at once, resulting in verification and correction circuits that are optimal in the number of required measurements and CNOTs.
The method is publicly available as an open-source tool as part of the \emph{Munich Quantum Toolkit}~(MQT)~\cite{willeMQTHandbookSummary2024} at \href{https://github.com/cda-tum/mqt-qecc}{https://github.com/cda-tum/mqt-qecc}, allowing fellow peers to create state preparation circuits for upcoming codes and codes not considered in this work.

For future work, it would be interesting to extend the proposed methods to codes beyond distance four, necessitating the correct handling of two or more \emph{independent} errors, i.e.,~also within the conditional correction circuits, requiring even more and potentially repeated measurements~\cite{peham2024automated,dennisTopologicalQuantumMemory2002}.

\section*{Acknowledgments}
\scriptsize
\noindent
This research is part of the Munich Quantum Valley, which is supported by the Bavarian state government with funds from the Hightech Agenda Bayern Plus and all authors acknowledge funding from the European Union’s Horizon Europe research and innovation programme under Grant Agreement No. 101114305 (“MILLENION-SGA1” EU Project). 

\noindent
L.S., T.P., L.B., and R.W. acknowledge funding from the European Research Council (ERC) under the European Union’s Horizon 2020 research and innovation program (grant agreement No.\ 101001318). 
\noindent
M.M. gratefully acknowledges support by the ERC Starting Grant QNets through Grant No. 804247 and support by the BMBF project MUNIQC-ATOMS (Grant No. 13N16070), by the Deutsche Forschungsgemeinschaft (DFG, German Research Foundation) under Germany’s Excellence Strategy “Cluster of Excellence Matter and Light for Quantum Computing (ML4Q) EXC 2004/1” 390534769.

\printbibliography

\end{document}